\documentclass[11pt]{article}
\title{Comment on ``Conceptual Inadequacy of Shannon Information ...'' by
C. Brukner and
A. Zeilinger}
\author{Michael J. W. Hall\\
Theoretical Physics, IAS\\Australian National University\\
Canberra ACT 0200, Australia}
\date{}
\begin{document}
\maketitle
%\newpage

\begin{abstract}
It is pointed out that the case for Shannon entropy and von Neumann
entropy,
as measures of uncertainty in quantum mechanics, is not as bleak as 
suggested in quant-ph/0006087.
The main argument of the latter is based on one particular interpretation of
Shannon's H-function
(related to consecutive measurements), and is shown explicitly to fail for
other physical interpretations.
Further, it is shown that Shannon and von Neumann entropies have in fact a
common fundamental
significance, connected to the existence of a unique geometric measure of
uncertainty for classical and quantum ensembles.
Some new properties of the ``total information measure" proposed in
quant-ph/0006087
are also given. 
\end{abstract}
\renewcommand{\thesection}{\Roman{section}}
\renewcommand{\thesubsection}{\Alph{subsection}}

\section{Interpretations of Shannon entropy}

Note that the term ``Shannon entropy'' will be used throughout for Shannon's $H$ function, rather than the term
``Shannon {\it information}'' used in [1], as the latter quantity in
general involves a {\it difference} of entropies \cite{ash}.

The main argument given in [1], against the use of Shannon entropy for
quantum measurements, relies
on showing that a particular interpretation of this quantity (involving
consecutive measurements)
does not accord with quantum mechanics. Here it is pointed out
that an alternative interpretation
may be given which does not involve consecutive measurements in any way.
Hence {\it no} general conclusions on the adequacy of
Shannon entropy can be drawn from the argument in \cite{bruk}.

Shannon showed that his $H$ function could be derived from a set of axioms
for ``uncertainty'' or
``randomness'' \cite{shan}. Numerous minor variations on these axioms have
since been used \cite{ash, behara},
and the discussion in Sec. III of [1] centres on the Faddeev form of
the so-called ``grouping axiom''
\cite{ash},
\begin{equation} \label{fadd}
H(p_1,p_2, ..., p_{n-1}, q_1, q_2) = H(p_1,p_2, ...,p_n) + p_n H(q_1/p_n,
q_2/p_n) ,
\end{equation}
for discrete probability distributions $(p_1, ...,p_n)$ and
$(q_1/p_n,q_2/p_n) $ (where $p_n=q_1+q_2$). This axiom, together with
axioms for the continuity and symmetry of
$H$ with respect to its arguments, leads uniquely to
\begin{equation} \label{shanent}
H(p_1,p_2,...,p_n) = -C \sum_i p_i \ln p_i ,
\end{equation}
where $C$ is an arbitrary multiplicative constant \cite{behara}.

Now, Brukner and Zeilinger induce their interpretation of Shannon entropy,
 in Sec.
III of [1], via an interpretation
of the axioms from which it is derived. Since the Faddeev form
(\ref{fadd}) of the grouping axiom is not physically
transparent, they introduce a physical justification for it 
based on consecutive
measurements
and their joint probabilities. 
They then reject this axiom (and consequently the 
Shannon entropy) as ``inadequate'' for
quantum measurements, essentially because classical joint probabilities do
not exist for noncommuting quantum observables.

However, the above argument immediately becomes inapplicable when an {\it
alternative} form of the grouping axiom, with a physical justification in
which consecutive measurements play no part, is used. In particular,
define two distributions to be 
{\it non-overlapping} if and only if there is some measurement which can
distinguish between them
with certainty. Thus if some outcome has a non-zero probability
of occurrence for one
such distribution, then it has a {\it zero} probability of occurrence for the
other. 

Suppose now that one prepares a mixture of 
non-overlapping distributions, each having its own ``randomness'' or
``uncertainty''. The randomness of outcomes is then expected to increase
on average,
due to the information thrown away by mixing, ie, due to the randomness arising from the mixing probabilities. The 
grouping axiom may then be
formulated as requiring this expected increase to be {\it additive} \cite{hall}:

{\it the randomness of a mixture of non-overlapping distributions is equal
to the average randomness of the individual distributions, plus the
randomness of the
mixing distribution}.

For example, note that the $(1/2, 1/3, 1/6)$ distribution discussed in Sec.
III of [1] is equivalent to an
equally weighted mixture of the two non-overlapping ensembles $(1, 0, 0)$ 
and
$(0, 2/3, 1/3)$. The
above form of the grouping axiom then implies that
\begin{equation} \label{example}
H(1/2, 1/3, 1/6) = \frac{1}{2} H(1,0,0) + \frac{1}{2} H(0,2/3,1/3) +
H(1/2,1/2)
\end{equation}
(where $H(1,0,0)$ is easily shown to vanish - see below). Further, the
Faddeev form of the grouping axiom in Eq.
(\ref{fadd}) is recovered by
decomposing the distribution $(p_1,p_2,.,p_{n-1},q_1,q_2)$ into the mixture
of non-overlapping
ensembles $(1,0,...,0,0,0)$, $(0,1,...,0,0,0)$, $...$, $(0,0,...,1,0,0)$,
$(0,0,...,0,q_1/p_n,q_2/p_n)$,
with respective mixing coefficients $p_1,p_2,$ $...,$ $p_n$, and noting that
$H(1,0,...,0,0,0)$ etc. must vanish
(this vanishing of uncertainty is of course expected for such distributions, and
follows immediately when the above form of the grouping axiom is combined
with the 
symmetry axiom for $H$, for the case of an 
equally weighted mixture of such distributions).

The grouping axiom thus has an alternative form with a physical
interpretation involving only the
notions of mixtures and non-overlapping ensembles, and having {\it no} reference
to consecutive measurements.
It follows that the argument in Sec. III of [1]
shows only that the {\it particular} interpretation of the Shannon
entropy given there, rather than the
Shannon entropy itself, is ``inadequate''.

\section{Geometric and operational significance of Shannon and von Neumann
entropies}

In Sec. IV of [1], Brukner and Zeilinger make the valid point that the
Shannon entropy of a quantum measurement is not invariant under unitary
transformations, and hence is not suitable as an invariant measure of
``uncertainty'' or ``randomness'' for quantum systems. They further note
that, for example, simply summing up Shannon entropies for three orthogonal
spin directions of a spin-1/2 particle does not provide an invariant
measure. However, they reject the von Neumann entropy as an appropriate
generalisation for the uncertainty of a quantum system, essentially on the
grounds that it is equal to a Shannon entropy only for the ``classical''
case of a measurement diagonal in the same basis as the density operator of
the system (although the same ``criticism'' holds for their 
proposed measures $I(p)$ and $I(\rho )$ in Eqs. (\ref{inf}) and 
(\ref{sum}) below). 

I wish to point out that
 there is in fact a very deep connection between Shannon and von
Neumann entropies as measures of uncertainty, connected 
to the existence of a {\it unique} measure of
uncertainty for classical and quantum systems with the geometric
properties of a ``volume''. 
In particular, consider a measure of the {\it
volume} (or {\it spread}) of a classical or quantum ensemble, which
satisfies:

(i) {\it the volume of any mixture of non-overlapping ensembles, each of
equal volume, is no greater than the sum of the component volumes (with
equality for an equally-weighted mixture)}

(ii) {\it the volume of an ensemble comprising two subsystems is no greater than the
product of the volumes of the subsystems (with equality when the
subsystems are uncorrelated)}

(iii) {\it the volume of an ensemble is invariant under all
measure-preserving transformations of the underlying space}.

These postulates are seen to be {\it independent} of whether the
ensemble is classical or quantum, and are discussed in detail in \cite{hall} 
(where the second
postulate is shown to 
 correspond to the Euclidean property that the product of
the lengths obtained by projecting a volume onto orthogonal axes is 
never less than the original volume). 
It is shown in \cite{hall}
that the only continuous measure of uncertainty $V$ which satisfies these
postulates is
\begin{equation} \label{vol}
V = K e^S ,
\end{equation}
where $K$ is a multiplicative constant, and $S$
denotes the Shannon entropy for
classical ensembles and the von Neumann entropy for quantum ensembles. It
is worth noting that this 
result also holds for the case of {\it continuous} classical distributions.

Thus the exponential of the entropy is a fundamental geometric measure of uncertainty
for both quantum and classical ensembles. Indeed,
volume may be taken as the primary physical quantity, and entropy then {\it defined}
as its logarithm. Note that this approach to entropy is very different to
the axioms discussed in Sec. I above, and leads to an {\it additive} 
rather than a
multiplicative constant for entropy. Further discussion and applications
may be found in \cite{hall}.

It is concluded that, in the context of uncertainty measures, the von Neumann entropy {\it is} in fact an
appropriate quantum generalisation of Shannon entropy. 

It is also perhaps worth
making some remarks on the ``operational'' significance of
von Neumann entropy, 
in the light of the discussion in [1]. Suppose that one makes
sufficient measurements on copies of a quantum ensemble to be able to
accurately estimate the density operator of the ensemble. For example, as
noted in [1], this may be done if the distributions of a sufficient number
of non-commuting observables are accurately determined (and is of course the
basis of quantum tomography). It follows that, having the density operator
at hand, one can immediately calculate the von Neumann entropy. The latter 
quantity thus 
has a perfectly good ``operational'' definition, and indeed
differs in this regard {\it no more and no less} from any functional of the
density operator, including the ``total information'' proposed in Sec. V of
[1] (see also Eq. (\ref{sum}) below).

\section{Some properties of ``total information''}

Brukner and Zeilinger propose a measure of ``total information'' for quantum
systems which has some interesting properties. Here I would like to point
out some further relationships, and also to demonstrate a property similar to the additivity property
found in [1], but which does {\it not} depend on the existence of a complete set of mutually 
complementary observables (which at present are only known to exist for Hilbert space dimensions that are prime or 
powers of 2).

First, let
\begin{equation} \label{inf}
I(p) = \sum_j (p_j - 1/n)^2 = \sum_j p_j^2 -1/n
\end{equation}
denote the information measure defined in Eq. (17) of [1] for distribution
$(p_1,...,p_n)$. If this distribution is generated by measurement of some
Hermitian observable $A$ on an $n$-dimensional Hilbert space, then it may 
alternatively be denoted by $I(A)$. Now, suppose that the Hilbert space admits a
complete set of mutually complementary observables, i.e., $n+1$ observables
$A_1, ..., A_{n+1}$ such that the distribution of any one observable is
uniform for an eigenstate of any other \cite{ivan}. It follows that one
has the general ``reconstruction'' property \cite{ivan}
\begin{equation} \label{ivan}
\rho = \sum_i \rho (A_i) - 1 ,
\end{equation}
where $\rho (A_i)$ denotes the density operator corresponding to a
projective measurement of $A_i$ on an ensemble described by $\rho$.
As shown in
Sec. V of [1], the additivity property 
\begin{equation} \label{sum}
\sum_i I(A_i) = {\rm tr} [(\rho - 1/n)^2] = {\rm tr}
 [\rho^2] -1/n =\colon I(\rho)
\end{equation}
then follows,
where $I(\rho )$ is the natural quantum 
generalisation of $I(p)$, called the ``total information'' [1]. Thus the quantum
information measure is just the {\it sum} of the classical information
measures, over a complete set of mutually complementary observables.

Eq. (\ref{sum}) is a very nice property relating the quantum and classical
contexts. Noting that the second term in each of Eqs. (\ref{inf}) and
(\ref{sum}) can be interpreted as the square of the ``distance'' between the
state of the system and a maximally-random state, this additivity property can
be
viewed as a type of Pythagorean connection between quantum and classical
distances. It may also be re-expressed as a relation between the quantum and
classical ``inverse participation ratios'' \cite{hall, inverse}
\begin{equation}
R(\rho )= [{\rm tr} \rho^2 ]^{-1}, \hspace{1cm}
R(p)=[\sum_j p_j^2 ]^{-1}
\end{equation}
(corresponding to non-Euclidean measures of ``volume'' \cite{hall}). 
In particular, Eq. (\ref{sum}) implies that 
\begin{equation} 
1/R(\rho ) = \sum_i 1/R(A_i) - 1 , 
\end{equation} 
which is formally analogous to the reconstruction property Eq.
(\ref{ivan}).

However, as noted in [1], the {\it existence} of $n+1$ mutually
complementary observables has in fact only been shown for the cases that $n$ is
prime or a power of 2. This puts the general applicability of Eq.
(\ref{sum}) in doubt. 

I wish to point out here that there is in fact a
similar relation between $I(\rho )$ and $I(p)$
which does {\it not} depend on the existence of a complete set of
complementary observables. 
In particular, instead of summing $I(A)$ over a specific group of
observables, one can instead {\it average} $I(A)$ over {\it all}
(non-degenerate Hermitian) observables. Such observables differ only by
unitary transformations, and if $dU$ denotes the normalised invariant Haar
measure over the group of unitary transformations $\{ U\}$, it can be shown
(see Appendix) that
\begin{equation} \label{int}
I(\rho) = (n+1) \int I(UAU^\dagger ) dU .
\end{equation}
Thus the quantum information 
measure is proportional to the average of the classical
measure, over all observables. 
This is clearly similar in spirit to Eq. (\ref{sum}),
which corresponds to replacing the 
average over all observables in Eq. (\ref{int}) by an
average over $n+1$ mutually complementary observables. The two averages are
thus equivalent for this information measure.
%\newpage

{\bf APPENDIX}

To prove Eq. (\ref{int}), note that for an observable $A$ with eigenstates
$\{ |a_j\rangle\}$ that
\begin{equation}
I(A) = \sum_j \langle a_j|\rho |a_j\rangle^2 - 1/n.
\end{equation}
For each term in the sum, the average over all observables 
is independent of $|a_j\rangle$
(since it is unitarily invariant), and hence $|a_j\rangle$ may be replaced
by a common state, $|a\rangle$ say, to give 
\begin{equation} \label{temp}
\int I(UAU^\dagger ) dU = n \int \langle a|\rho |a\rangle^2 d\Omega_a -
1/n,
\end{equation}
where $d\Omega_a$ is the normalised invariant measure over pure states
\cite{skyora}.
By expanding $\rho$ into eigenstates, and again noting the unitary
invariance of the average, it follows that
the righthand side of this equation has the form $\alpha {\rm tr} [\rho^2] +
\beta$, where $\alpha$ and $\beta$ 
are definite integrals. Further, noting that the
righthand side vanishes for the maximally mixed state $\rho=(1/n){\bf 1}$,
one must have $\beta =-\alpha /n$.  Hence 
\begin{equation} \label{cruc}
\int I(UAU^\dagger ) dU = \alpha I(\rho ).
\end{equation}

Finally, to determine $\alpha$, assume that $\rho$ corresponds to some pure
state $|b\rangle$. Hence $I(\rho )=(n-1)/n$ from
Eq. (\ref{sum}), and substitution in Eqs.
(\ref{temp}) and (\ref{cruc}) yields
\begin{equation}
\alpha = \left[ n^2 \int |\langle a |b\rangle |^4 d\Omega_a - 1
\right] /(n-1).
\end{equation}
The integral may be evaluated either via Eq. (38) of \cite{hallpla} (which
gives the probability distribution for the 
variable $Y=|\langle a|b\rangle |^2$), or 
via the more general method in Appendix A of
\cite{jones}, as $2/n/(n+1)$, and Eq. (\ref{int}) immediately follows.
%\newpage

\end{document}